\documentstyle[epsfig]{elsart}

\begin{document}
\begin{frontmatter}
\title{
First results on characterization of Cerenkov
images through combined use of Hillas, fractal
and wavelet parameters}
\author[KARLSRUHE]{A.~Haungs}
\author[BOMBAY]{A.K.~Razdan}
\author[BOMBAY]{C.L.~Bhat}
\author[BOMBAY]{R.C.~Rannot}
\author[KARLSRUHE]{H.~Rebel}
\address[KARLSRUHE]{Forschungszentrum Karlsruhe, Institut f\"ur 
Kernphysik III, Postfach 3640, D-76021 Karlsruhe, Germany}
\address[BOMBAY]{Bhabha Atomic Research Centre, Nuclear Research 
Laboratory, \\ Mumbai - 400 085, India} 

\begin{abstract}
Based on Monte Carlo simulations using the CORSIKA code, it is
shown that Cerenkov images produced  by  ultrahigh energy $\gamma$-rays
and cosmic ray nuclei (proton, Neon and Iron) are fractal in nature.
The resulting multifractal and wavelet moments when employed in association
with the conventional Hillas parameters as inputs to a properly-trained
artificial neural network are found to provide more efficient primary
characterization scheme than the one based on the use of Hillas or
fractal parameters alone.
\end{abstract}

\end{frontmatter}

\vskip 0.5cm
\section{Introduction}
Recent developments on the simulation and experimental
fronts in Cerenkov Imaging Technique (CIT) in the field of TeV
$\gamma$-ray astronomy have made it possible to preferentially pick
up atmospheric Cerenkov events from $\gamma$-ray progenitors and
substantially reject cosmic-ray-generated background events
($\gamma$-ACE and C-ACE respectively, hereonwards) and thereby 
significantly augment the detection sensitivity of ground-based, 
very high energy (VHE) $\gamma$-ray telescopes \cite{fegan}. 
For this purpose, a 2-dimensional Cerenkov image, as recorded by 
a fast photomultiplier tube (PMT)-based,
multi-pixel camera, placed in the focal-plane of a large light
collector ($\ge 10\,$m$^2$), is subjected to some
pre-processing routines like 'flat-fielding' and 'cleaning' and is 
then parameterized as per the second- and higher-moments 
prescription proposed first by Hillas \cite{hillas}. 
The resulting Hillas parameters, superceded by the supercuts 
image processing strategy, have helped to reject the C-ACE
at $\ge99.7\% $ level, while permitting to retain the signal events
($\gamma$-ACE) at $\sim 50\%$ level, leading, in turn, to a typical
quality factor, Q of $\sim10$ for the prototype Whipple imaging
Cerenkov telescope \cite{whipple}. 
As this value appears to be close to the 
limiting $Q$ that can be achieved through the use of 
moment-fitting procedures to the real or simulated Cerenkov images, 
the problem that warrants a serious attention now is whether a 
fundamentally different (and perhaps more primordial) 
image-processing approach can be found which will, on its own, 
or in conjunction with the presently-used
image-analysis technique, help to classify the ACE progenitor
more efficiently and thereby be able, not only to segregate 
$\gamma$- and C-ACE but also various primary cosmic-ray nuclear 
groups (different atomic number, Z, ranges), with respect to 
one another. 
While, on the one hand, this should enable the imaging 
Cerenkov telescopes to pick-up
weaker $\gamma$-ray signals more quickly, on the other hand, 
it may also permit these instruments to function as a 
low-resolution cosmic-ray spectrometer, at least, to the extent 
of differentiating various cosmic-ray particles into low Z  
(Hydrogen or proton-like), medium Z (Neon-like) 
and high Z (Iron-like) nuclei. Thus, apart from performing
$\gamma$-ray astronomy investigations, this may hopefully
pave the way for deploying imaging Cerenkov telescopes eventually
in a supplementary mode of operation for independent,
cosmic-ray mass-composition studies in the ultra-high energy 
region -- an important, outstanding problem in its own right, 
which is currently engaging lot of attention internationally.  
Motivated by this consideration,
we investigate here the possibility of subjecting
the Cerenkov image data to perhaps more general, albeit as-yet 
untried, multifractal and wavelet analyses in anticipation of 
deriving independent parameters which can supplement 
presently-in-use Hillas moments for event-characterization purposes. 
The paper discusses here first results from these
exploratory studies, based on Cerenkov image data simulated for the
Imaging Element of the 4-element TACTIC array, using the
CORSIKA air-shower code.

\section{TACTIC}
The acronym TACTIC stands for a {\bf T}eV {\bf
A}tmospheric {\bf C}erenkov {\bf T}elescope with an {\bf I}maging
{\bf C}amera \cite{tactic}. 
Keeping in mind the scientific merits of adopting the
above-referred dual-purpose detection strategy,
the TACTIC has been specifically designed,
on one hand, to carry out high-sensitivity
spectral and temporal investigations on VHE $\gamma$-ray sources on
clear, dark nights and, on the other, to utilize it effectively
\begin{figure}[ht]
\begin{center}
\hspace*{-0.5cm}
\epsfig{file=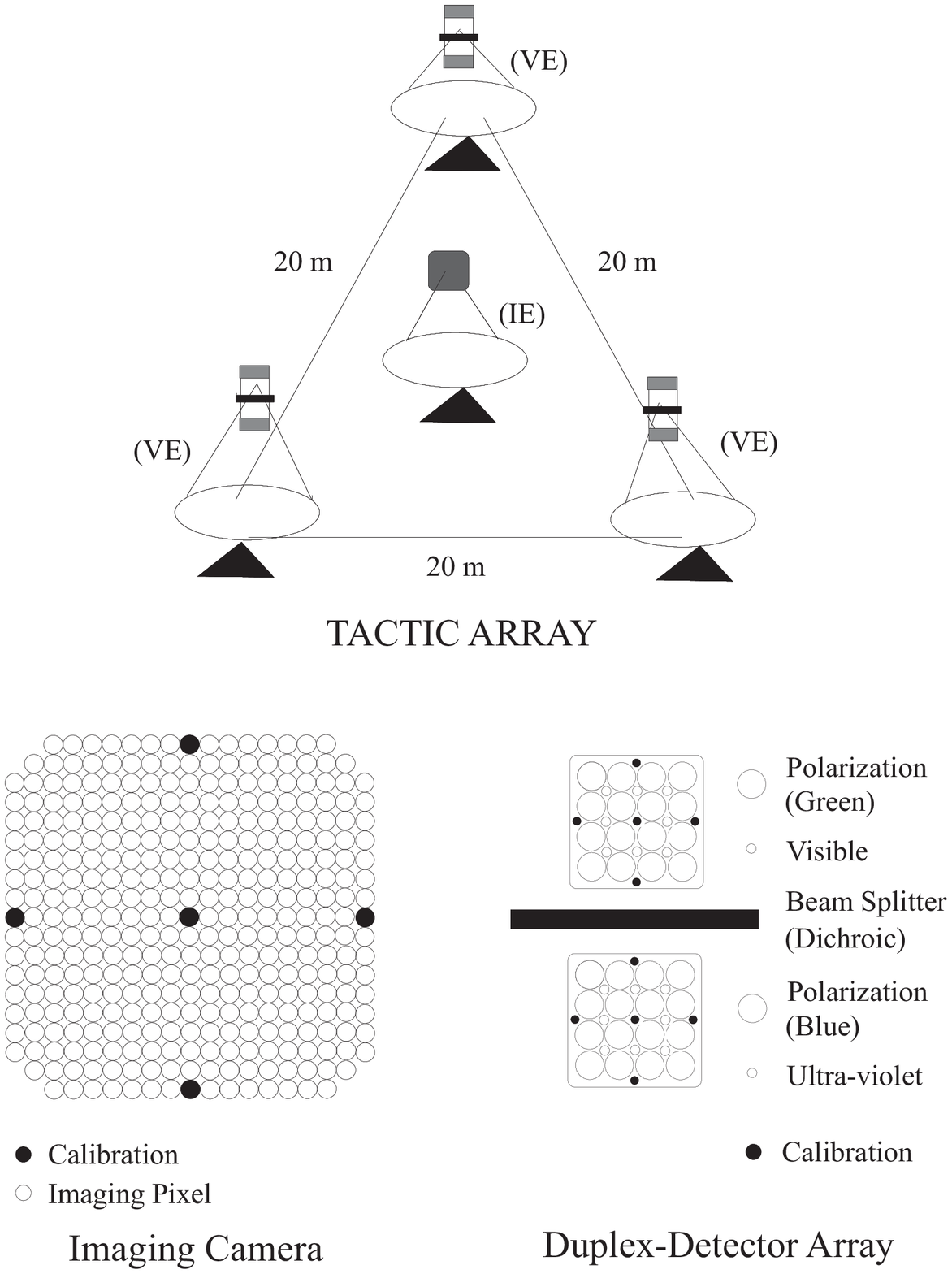,width=12.5cm}
\caption{\it Sketch of the TACTIC array and the cameras, presently under installation, for high-sensitivity spectral and temporal investigations on
$\gamma$-ray sources in the photon energy range of $\sim 0.5-10\,$TeV 
(details see text).
The array would also investigate UHE cosmic-ray elemental composition
(50-500 TeV per particle) in a supplementary mode of operations, going
to be employed under partially moonlit conditions when normal 
$\gamma$-ray sources studied are not conventionally performed.}
\label{fig1}
\end{center}
\end{figure}
during semi-lit portions of a night (normally, shut-down period for
atmospheric Cerenkov systems) for cosmic-ray mass-composition
investigations in 10's-100's TeV particle energy region.
This instrument consists of an array of 4 atmospheric
Cerenkov elements each using a tessellated optical collector of
$9.5\,$m$^2$ light collection area and a synchronized,
computer-controlled alt-azimuth drive system. As is evident from
Fig.\ref{fig1}, the 创Imaging Element创 (IE) of this compact 
telescope array is located at the centroid of an equilateral 
triangle of $20\,$m-side and the 3 创Vertex Elements创 (VE) are placed 
at the vertices of this triangular configuration. 
The individual $60\,$cm-diameter mirror facets
of the light collector are mounted so as to follow a
quasi-Davis-Cotton surface profile, leading to an 
on-axis spot-size of $\sim <15\,$am (FWHM).  
The IE has a fast photomultiplier tube (PMT)-based
349-pixel Imaging Camera in its focal-plane, covering a
field-of-view (FoV) of $\sim 6^\circ \times 6^\circ$ truncated square
with a pixel resolution of $\sim 0.31^\circ$ diameter. 
The 3 VE's, on the other hand, are provided with somewhat 
unconventional focal-plane instrumentation which consists of 
duplex PMT arrays of assorted sizes, 
placed across a beam-splitter/dichroic sheet
assembly, as shown in Fig.\ref{fig1}~.  
The resulting non-image parameters of the VE's 
are planned to be deployed 
in conjunction with the high-definition imaging data provided by
the IE for a more efficient characterization of the recorded ACE in
relation to the nature of the progenitor particles.
The IE of the TACTIC is presently operational with a 81-pixel (9
$\times$ 9) camera and it is being regularly used since March 1997
for $\gamma$-ray source observations \cite{bhat}.  
The 349-pixel IE of the TACTIC array is scheduled to become 
operational by December 1998.  
In anticipation of this commissioning schedule,
comprehensive Monte Carlo simulation studies are presently underway
to provide specific guidelines for optimizing event characterization
strategies and thereby enable this instrument to carry out 
the above-outlined VHE $\gamma$-ray astronomy and
UHE cosmic-ray mass-composition investigations through the
atmospheric Cerenkov detection route.  
Here, in the first report on this work, we first establish that 
Cerenkov images recorded by an instrument like the TACTIC-IE
for $\gamma$-ray, proton, Neon and Iron progenitors
have a fractal structure, 
and then go on to demonstrate the feasibility of segregating 
these event families comparatively more efficiently by using 
Hillas parameters in association with a selection of 
multifractal dimensions and wavelet moments.

\section{CORSIKA-based Cerenkov Image data-bases}
The data-bases for carrying out the present
feasibility-demonstration studies were generated using the CORSIKA
(Version 4.5) air-shower code \cite{capdevielle,heck} 
with Cerenkov option, and the use of the high energy interaction
model VENUS \cite{venus} and the model GHEISHA \cite{gheisha}
for interactions at lower energies 
(E$_{lab} < 80\,$GeV).
A rectangular matrix of 60 $\times$ 4
detector elements, each $5\,$m $\times$ $5\,$m in dimensions, is
folded into the CORSIKA simulation programme for this purpose,
in conformity with the actual geometrical configuration of the TACTIC
array and the size of its 4 light-collectors \cite{rannot}.
\begin{figure}[htb]
\begin{center}
\hspace*{-0.5cm}
\epsfig{file=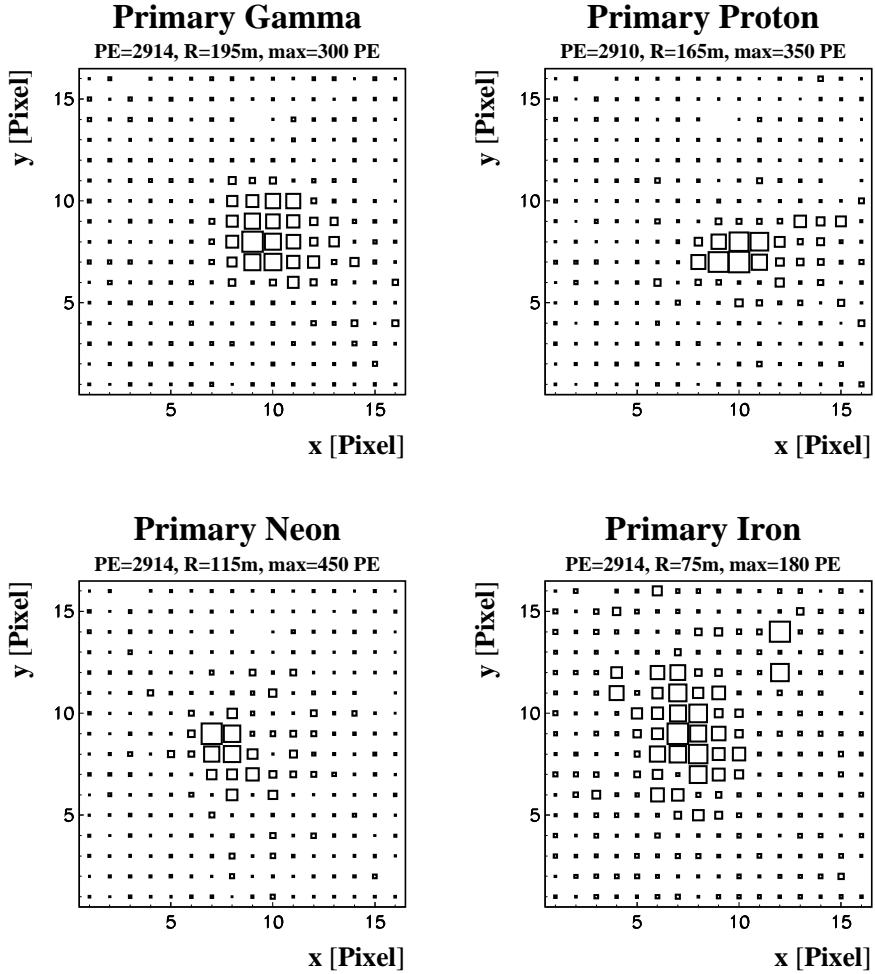,width=13.cm}
\caption{\it Examples of simulated Cerenkov images for a 16 x 16
pixel camera for different progenitors with approximately 
equal total number of photoelectrons (PE). The box size represents
the number of PE per pixel, the number of PE for the largest pixel 
(max) is indicated as well as the distance R of the camera from 
the shower axis.}
\label{fig2}
\end{center}
\end{figure}
The Cerenkov data bases of 100 showers each for $\gamma$-rays, proton and
Neon and Iron nuclei have been generated. These simulated data bases are
valid for the altitude (1700 m) and magnetic field values of
Gurushikhar, Mt. Abu, the permanent location of the TACTIC in the Western
Indian state of Rajasthan. The zenith angle $\Theta$ of the primary is 
fixed at $40^\circ$ for gamma and $40^\circ\pm2^\circ$ 
for protons, Neon and Iron,
larger $\Theta$ being preferred to achieve twin benefits
of a higher primary threshold energy and a larger effection 
collection area.
$\gamma$-rays of energy $50\,$TeV and protons, Neon
and Iron nuclei of $100$ TeV energy (the factor of two higher 
energy for nuclear-progenitors being chosen to have roughly 
comparable average Cerenkov photon densities in all the 4 cases) 
have been considered.
While $\gamma$-ray primaries are supposed to be incident along 
the principal axis of the TACTIC IE (as expected for $\gamma$-rays 
from a point-source) the 3 types of the cosmic-ray progenitors 
(protons, Neon and Iron) have their angles of
incidence randomly oriented in a circular field of view of
$3^\circ$ radius around the IE axis (in accord with the well-known
isotropic angular distribution of cosmic rays).
To keep computer time and the Cerenkov photon file size within 
manageable limits for a given event, the Cerenkov photons are 
generated in the restricted wavelength region, 
$\lambda \sim (300 - 320)\,$nm and simulation run
bunch size has been fixed at 20~.
Cerenkov photons likely to be received at a
given element with $\lambda$ outside the above-referred sample 
window are generated off-line, using the well-known Cerenkov 
radiation spectral law $\sim \lambda^{-2}$.  
Other exercises done subsequently in the off-line mode include 
(i) taking proper account of the $\lambda$-dependent atmospheric 
extinction suffered by the individual Cerenkov photons emitted in 
the overall wavelength interval $\lambda \sim (300-600)\,$nm, 
(ii) ray-tracing Cerenkov photons, incident on
each $60\,$cm-diameter facet of the IE mirror into the
focal-plane of the light receiver, and 
(iii) deriving the number of photo-electrons (PE) likely to be 
registered by each of the 349 photomultiplier (PMT) pixels of the 
IE camera after accounting for the reflection coefficient of 
the mirror and the quantum efficiency of the PMT pixels. 
Fig.\ref{fig2} gives representative examples of the
images thus generated for the IE in response to 4 progenitor types
($\gamma$-particles, proton, Neon and Iron nuclei).
In anticipation of the requirements of the associated analysis 
procedure, only the square-grid, comprising the innermost 
16 x 16 pixels of the TACTIC imaging camera, has been considered.  
Each of these 256 pixels, (which also include the pixels with 
information on the simulated Cerenkov image) have been injected 
with a photomultiplier noise component $\sim 4\,$PE. The noise 
injected in the image follows Poissonian distribution.  
For each simulated event, there exist
images corresponding to as many detector locations as folded
to the CORSIKA output.

\section{Classification Schemes}
\subsection{Hillas Image Parameters}
The Cerenkov image recorded by a TACTIC-like imaging telescope
represents the 2-dimensional distribution of the light pattern
produced in the terrestrial atmosphere following the incidence 
therein of a VHE/UHE $\gamma$-ray or a cosmic-ray primary.  
This image has embedded in it signatures which relate to the 
details of interactions which the progenitor particle as well 
its secondary and higher-generation by-product particles undergo 
in the atmosphere.  
\begin{figure}[htb]
\begin{center}
\epsfig{file=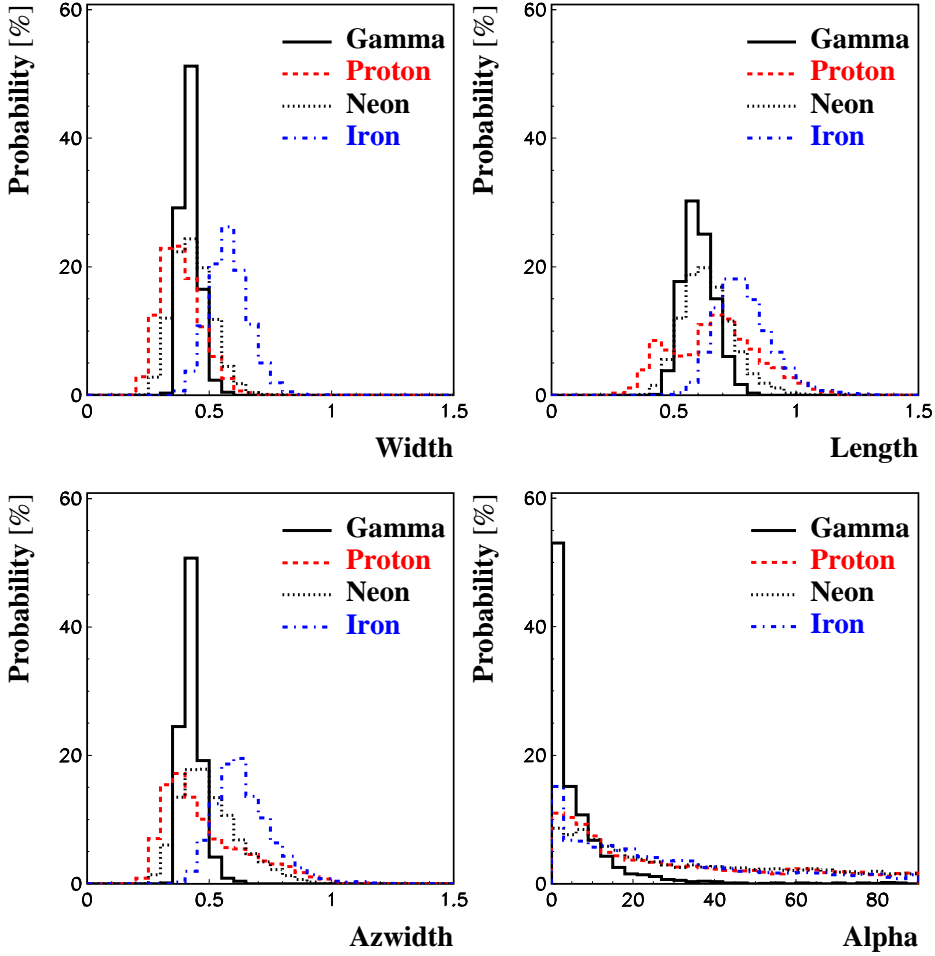,width=13.cm}
\caption{\it Distributions of Hillas parameters for all simulated
images in the range of PE from 1800 to 3000 for the different
progenitor particles of the ACE.}
\label{fig3}
\end{center}
\end{figure}
If properly interpreted, these signatures can be utilised to reveal
the nature and the energy of the progenitor particle.  
Starting off with the pioneering work of Hillas, the effort so 
far has concentrated on approximating the image to a geometrical 
ellipse and parameterizing it in terms of some characteristic
features like image 'shape' and 'orientation' with respect to a 
Cartesian frame of reference centred on the projection of the 
telescope axis on the image plane.  
Using the standard moments-fitting procedure \cite{fegan}, 
the typical parameters calculated for each image ellipse are
Length (L), Width (W), Azwidth (A), Miss (M), Alpha $(\alpha)$,
Distance (D) as per the known formulae.
Briefly stating here, the image Length and Width (shape parameter) 
refer to the dimensions of the major and minor axes of the 
projected ellipse and are related to the longitudinal and transverse
developments of the associated Cerenkov radiation pulse.
Similarly, the image 'orientation' parameter ($\alpha$) gives a 
measure of the angular displacement of the major axis of the 
image from the line connecting the image centroid to the telescope 
axis. 
For a $\gamma$-ray event from an on-axis point source, 
$\alpha$ is expected to be small, in practice generally
$\le 10^\circ - 15^\circ$, while for a cosmic-ray proton, 
all values of $\alpha$ are equi-probable, 
in recognition of the isotropic angular distribution of 
cosmic-ray events.
The results, obtained here for the IE of the TACTIC, 
using 4-component ($\gamma$, p, Ne and Fe) data-base, 
simulated as out-lined in section 3, are summarized 
in Fig.\ref{fig3}~.  
As expected, all the plotted parameters are
found to vary over a narrow range of values in the case of
$\gamma$-rays as against the corresponding situation for the nuclear
species because there exists a nuclear active channel where
large transverse momenta may be transferred to the pion 
secondaries, leading to a relatively broader Cerenkov image.
Among the latter events, Fe images are found to have the
narrowest distribution for most of the plotted parameters,
which is a reflection of the fact that, for
the same primary energy, an Fe shower reaches its peak growth at a
higher altitude than a corresponding EAS induced by a lower Z
primary, including a proton.  
However, it is evident from Fig.\ref{fig3} that there is a 
fairly high degree of overlap among the corresponding 
distribution profiles for an imaging parameter for these 
representative members of the cosmic-ray beam.  
Thus, speaking even qualitatively, it is apparent that, 
while Hillas image parameters can segregate $\gamma$-ACE from the 
C-ACE, they are not sufficiently sensitive to differences in Z 
and hence may not separate various nuclear progenitor types.

\subsection{Multifractal and Wavelet Moments}
Pattern-analysis procedures, based
on calculations of multifractal and wavelet moments of a structure,
have started becoming popular now, for they are more holistic 
and permit a more detailed examination of the morphology of a 
structure on different length scales.  
These analysis techniques provide a set of robust classifiers 
which are more sensitive to the structure
details and emphasize physical differences as against
statistical fluctuations. 
Guided by these considerations, we show
below that the present image data-base is amenable to the fractal
and the wavelet treatments and, as an important consequence, it is
possible to derive a set of independent parameters (multifractal
and wavelet moments) with better potential for characterizing
4-component data-base used for illustration in the present work. \\

Fractals are structures which display a self similar behavior
and fractal nature is quantitatively characterized by fractal
dimension \cite{mandelbrot}.  
It is possible to calculate multifractal moments
which quantify structures of multidimensional density 
distributions \cite{aharony}.
In anticipation of the requirements of the associated analysis 
procedure, only the square-grid, comprising the innermost 
16 x 16 pixels of the TACTIC Imaging camera, has been considered.  
Noise has been added to check the robustness of these image 
processing techniques against deleterious effects by noise 
contamination.  
Only images with a total number of $\ge$ 1800 PE have been used.  
We have calculated multifractal moments of each
simulated Cerenkov image by dividing the image into M = 4, 16, 64 
and 256 equally sized parts and by calculating the number of 
photoelectrons in each part.
M is related to the fractal scale-length $\nu$ by
$ M = 2^{\nu}\,$.
The multifractal moments given by the following expression 
have been computed:
\begin{equation}
G_q (M) = \sum_{j=1}^{M} (\frac{k_j}{N})^{q} \,\, , \,\, 
N \ne 0 \,\, ,
\end{equation}

where $N$ is the total number of PE in the image, $k_j$ is
the number of PE in the $j^{th}$ cell
and q is the order of the fractal moment.
If the Cerenkov image exhibits a self similar behavior the 
fractals moments $G_q$ show a power law
relation of the parameter of the length scale M:

\begin{equation}
G_q \propto M^{\tau_q} \,\, .
\end{equation}

The exponent $\tau_q$ is determined from $G_q$ by using the 
formula given below \\

\begin{equation}
\tau_q =  \frac{1}{\ln{2}} \frac{d{\ln{G_q}}}{d\nu} \,\, .
\end{equation}

The slope has been obtained from $\nu=1$ to $\nu=4$.
For a fractal structure, there exists a
linear relationship between the natural logarithm of the 
multifractal moment ($G_q$) and the fractal scale-length 
$\nu$ and the slope of this line, $\tau_q$ can be shown to be 
related to the generalized multifractal dimensions, $D_q$ by

\begin{equation}
\ D_q = \frac{\tau_q}{q-1} \,\, , \,\, q \ne 1 \,\, ,
\end{equation}

where q is the order of the moment and varies over the range
-6 $\le$ q $\le$ 6. For $q = 1$, $D_1$ is defined as equal to one.%
\begin{figure}[b]
\begin{center}
\epsfig{file=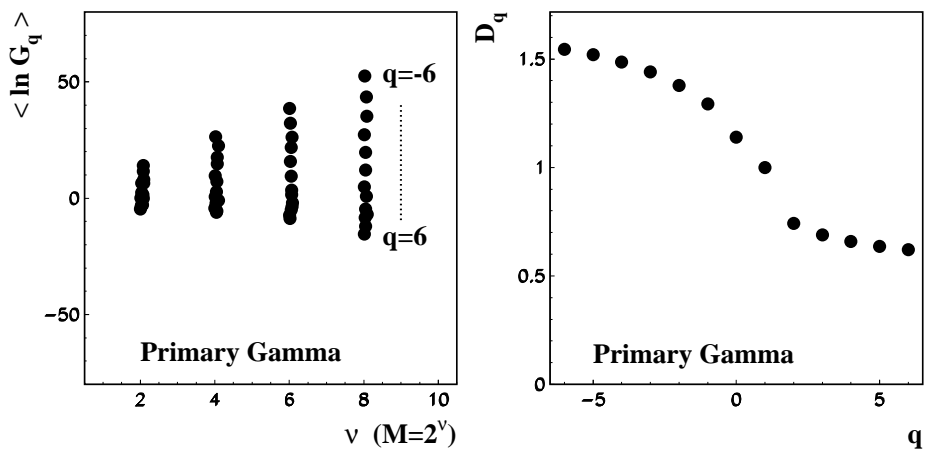,width=13.cm}
\caption{\it Multifractal moments vs. length-scale $\nu$ (left) 
and generalized multifractal dimensions vs. q (right) for one 
example of a $\gamma$-induced Cerenkov image displayed in 
Fig.\ref{fig2}~.}
\label{fig4}
\end{center}
\end{figure}
\begin{figure}[htb]
\begin{center}
\epsfig{file=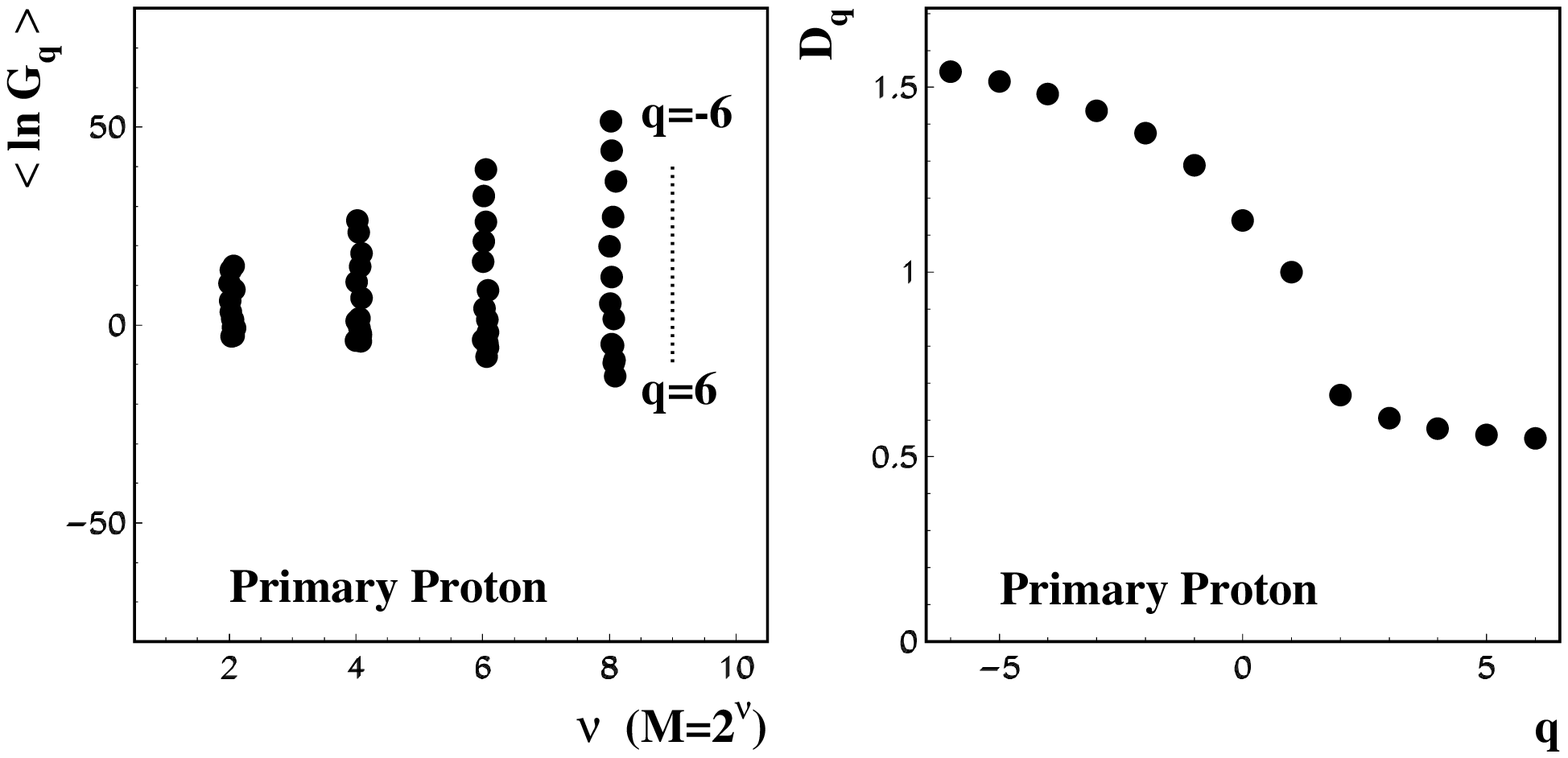,width=13.cm}
\caption{\it Multifractal moments vs. length-scale $\nu$ (left) 
and generalized multifractal dimensions vs. q (right) for one 
example of a proton-induced Cerenkov image displayed in 
Fig.\ref{fig2}~.}
\label{fig5}
\end{center}
\end{figure}
\begin{figure}[htb]
\begin{center}
\epsfig{file=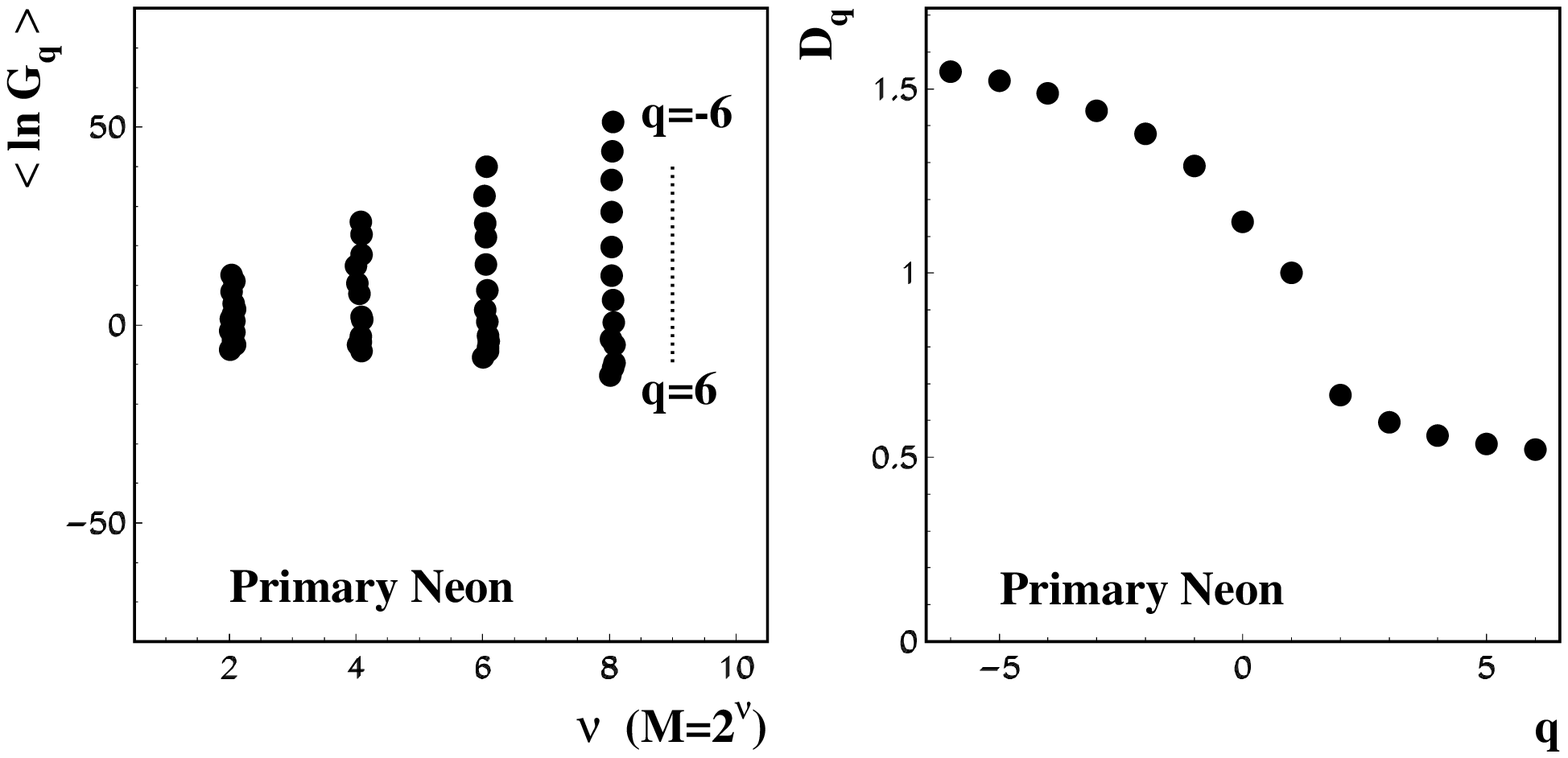,width=13.cm}
\caption{\it Multifractal moments vs. length-scale $\nu$ (left) 
and generalized multifractal dimensions vs. q (right) for one 
example of a Neon-induced Cerenkov image displayed in 
Fig.\ref{fig2}~.}
\label{fig6}
\end{center}
\end{figure}
\begin{figure}[htb]
\begin{center}
\epsfig{file=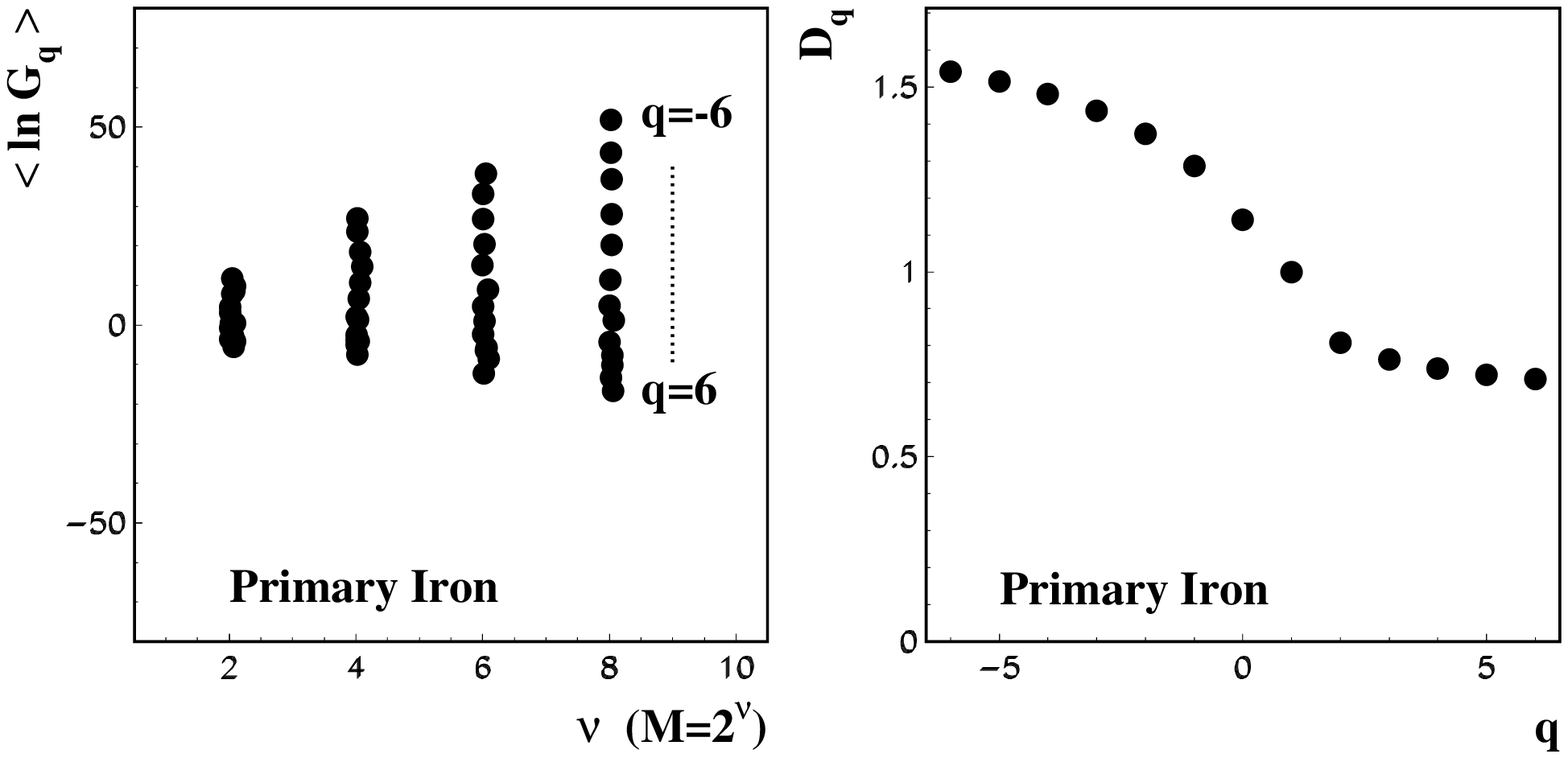,width=13.cm}
\caption{\it Multifractal moments vs. length-scale $\nu$ (left) 
and generalized multifractal dimensions vs. q (right) for one 
example of a Iron-induced Cerenkov image displayed in 
Fig.\ref{fig2}~.}
\label{fig7}
\end{center}
\end{figure}
For purposes of illustration, the results of this analysis are shown
in Figs.\ref{fig4}-\ref{fig7} for one example each of $\gamma$-ray, 
proton, Neon and Iron events respectively, simulated as per the 
details given in section 3.
The images of these 4 events, alongwith the injected
noise, are shown in Fig.\ref{fig2},
while the right and the left panels of Figs.\ref{fig4}-\ref{fig7} 
respectively present plots
of $G_q$ vs. $\nu$ and $D_q$ vs. q for orders of the moments
in the range $-6 < q < 6$. 
The power law relation of $G_q$ vs. $M$ is given in all images
(i.e. the pattern has a self similar behavior) and therefore
using the method of multifractal moments as parametrisation 
of the pattern is adaptable.
This result is significant in that 
they establish that Cerenkov images produced by
the 4 progenitor species, considered here, behave as fractals and as
such their structures are amenable for analysis through use of
multifractal dimensions $D_q$. 
A larger number of pixels at the camera would increase the usable
length scale ($\nu$) and therefore accuracy, 
but would not change the exponents $\tau$ of the pattern. 
Another fractal feature
is the saturation effected noted in the lower panels in the value of
$D_q$ with increasing q, thereby reassuring that the range 
$-6 < q < 6$, covered here is quite adequate.
The value of $D_{q_{max}}$ characterizes the location and the size 
of the largest irregularity in the image structure.  
This means that, more regular the image, the closer is $D_6$ 
to 1.  
On the other hand, the value of $D_{q_{min}}$ refer to image zones 
with lower photoelectron density than the overall image size
\cite{haungs}.  
Due to the fact that, in the present case, the Cerenkov image 
generally does not fill the entire camera FoV of
$6^\circ \times 6^\circ$ and the sizes of the ellipses do not 
vary over a large range, values of $D_q$ $<$ 0 are not useful in the
present application.  
$D_2$ is the so called correlation dimension \cite{hentschel},
which is widely used in analysing experimental pattern 
distributions in terms of fractal dimensions. %
\begin{figure}[htb]
\begin{center}
\epsfig{file=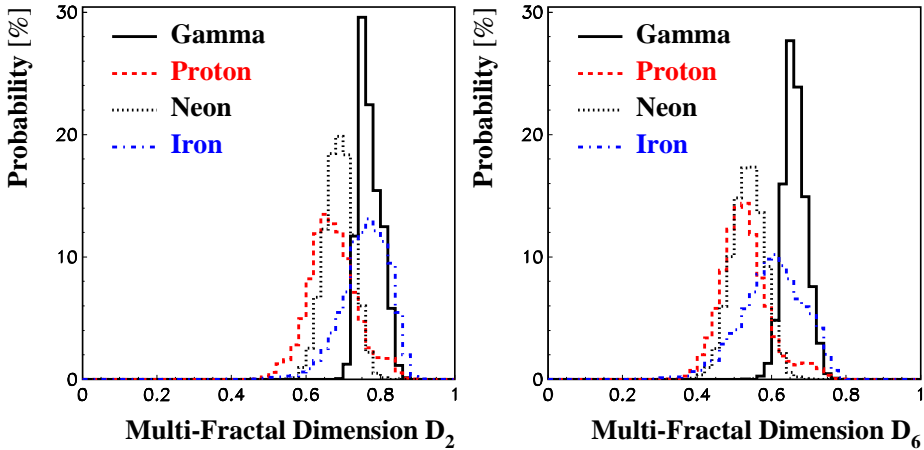,width=13.cm}
\caption{\it Distributions of the multifractal dimensions 
$D_2$ and $D_6$ for all simulated images in the range 
of PE from 1800 to 3000 for the different progenitor 
particles of the ACE.}
\label{fig8}
\end{center}
\end{figure}
In Fig.\ref{fig8}, we compare the distributions of the two multifractal 
parameters $D_2$ and $D_6$ for the four progenitor species 
comprising the data-base used in the present work.  
Both families of the plotted curves indicate that the peak
values of these two parameters are the smallest for protons and the
largest values for $\gamma$-rays, with Neon followed by Iron
images having in-between values.  
This is essentially opposite to the
behavior of the Length and Width Hillas parameters as is noted 
for these progenitor species in Fig.\ref{fig2}.  
This is so because, while these two Hillas
parameters are related to the image shape, the multifractal
parameters $D_2$ and $D_6$ reflect the overall regularity of 
the image structure.  
As $\gamma$-ray images produce the most regular images
amongst the four particle types considered here, $D_2$ and $D_6$ have
the largest peak values for $\gamma$-ray progenitors.  
Iron images are more regular than proton images, since, for the 
same total energy per nucleus, iron events have lesser energy 
per nucleon.  
This results in a smaller interaction length for iron primaries 
(as in the case of $\gamma$-rays) and hence more secondaries 
with lesser energy for particle  than what is expected in the case of
proton events.  
This leads to destroy a visible hadronic core in
the image of an Iron progenitor, while it survives in the case of a
proton image.  
Using  Hillas parameters $\gamma$-ray and proton images have a 
significant overlap in the corresponding distribution profiles, 
while Iron images are better segregated.  
Therefore, it follows that, by choosing a judicious
combination of the Hillas and multifractal parameters, it should be
possible to seek a better classification of these progenitor types. \\

By using wavelet analysis \cite{debauchies}, which can be
regarded as a sequence of versatile filtering processes which allows
to examine an image for presence of local structures on different
scale-lengths it has been possible to improve upon the results.
Wavelet and wavelet transforms may be viewed as generalizations
of the orthogonal Fourier transform.
The wavelet analysis technique does not use a series of sine 
and cosine functions, as employed in the Fourier analysis, 
but a more localized functions, called wavelets, 
(e.g., Haar-wavelet).  
As such, the wavelets can detect both the location and the scale 
of a structure in an image.  
The wavelets are parameterized by a scale (dilation
parameter) $a>0$ and a translation parameter) $b$ 
($-\infty <b<+\infty$) such that
\begin {equation}
\phi(x) = \frac{\psi (x-b)}{a} \,\, .
\end{equation}
The wavelet domain of one-dimensional function is rather two 
dimensional in nature; one dimension corresponding to the 
scale (a) and other corresponding to the space (translation b).  
Since we are analyzing a fractal; it is scaler
rather than the translation that is of interest to us.
When applied in the context of the TACTIC images, the wavelet
moment \cite{greiner} $W_q$ is given by:
\begin{equation}
W_q (M) = \sum_{j=1}^{M} (\frac{ | k_{j+1} - k_{j} | }{N})^{q}
\end{equation}
$k_j$ is the number of PE in the $j^{th}$ cell in a particular scale,
and $k_{j+1}$ in the $j^{th}$ cell in the consecutive scale.
\begin{figure}[t]
\begin{center}
\epsfig{file=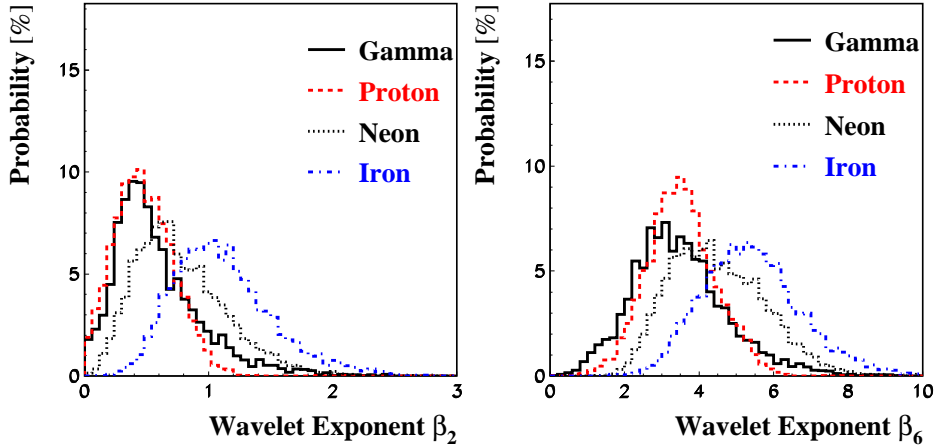,width=13.cm}
\caption{\it Distributions of the wavelet parameters $\beta_2$
and $\beta_6$ for all simulated images in the range of 
PE from 1800 to 3000 for the different progenitor 
particles of the ACE.}
\label{fig9}
\end{center}
\end{figure}
The wavelet moments have been obtained by 
dividing the Cerenkov
image into $M= 4, 16, 64, 256$ equally sized parts 
(with 64, 16, 4, 1 PMT, respectively) and counting the
number of photoelectrons in each part. The difference of
probability in each scale gives the wavelet moment.
Again a proportionality
\begin{equation}
W_q \propto M^{\beta_q}
\end{equation}
is given for the electron distribution in the Cerenkov images.
The two wavelet dimensions which we have
used for examining the structures of the Cerenkov images are 
the slopes $\beta$ of the best-fit regression line for the double
logarithmic distribution $W_q$ vs. M for $q=2$ and $q=6$ 
obtained for each image.
Fig.\ref{fig9}, gives the distribution of these two wavelet 
parameters $\beta_2, \beta_6$ for the simulated data-bases 
and belonging to the 4 progenitor species used in this work: 
$\gamma$-rays, protons and Neon and Iron nuclei.  
The steepness of the slope is found to gradually increase 
from gamma-rays to protons, followed by Neon
and Iron nuclei.  
It is well known that wavelet moments are sensitive to differences 
in the average numbers of photoelectrons in neighboring pixels 
on different length scales.  
Lesser this difference, i.e., more regular the image, the
flatter is the slope of the best-fit ($ ln W_q$ vs. $ln M$) 
regression line.
As Fe and Ne events are associated with a relatively larger number of
muons compared with proton (and $\gamma$-ray) events, the Cerenkov
images produced by these high Z nuclei are characterized by
local intensity peaks, resulting in higher values for the slope
parameter in case of high Z nuclei compared with protons (and
$\gamma$-rays).  This feature, which may be present in the 
structure of a Cerenkov image, is not readily exploitable 
through the conventional route, i.e., the use of Hillas parameters.  
This underlines the possible role of the wavelet parameters is 
being used as a supplementary set of parameters for better 
event-characterization purposes.

\subsection{Artificial Neural Network-based Classification}
We use the well known pattern recognition
capabilities of an artificial neural network (ANN), in order
to display the degree
by which the event-classification potential of the Hillas parameters
can be increased through the supplementary use of the 
fractal and wavelet parameters.  
From a statistical modeling point of view, the ANN technique 
represents a non-parametric event classification scheme.  
As it contains more free parameters, it requires relatively 
more training data.  
\begin{figure}[htb]
\begin{center}
\epsfig{file=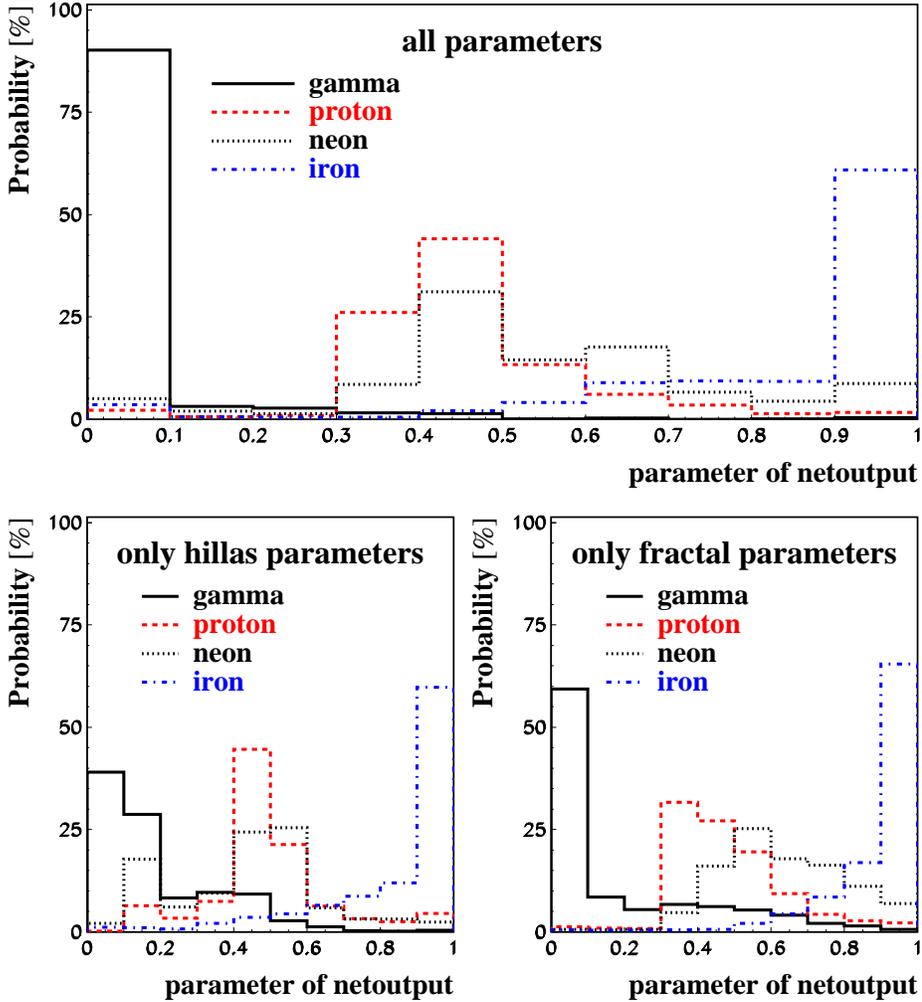,width=13.cm}
\caption{\it Distributions of the different primaries
in the output parameter of neural nets
with different sets of parameters. The nets are trained
with an independent event sample. Demanded output values are
0., 0.33, 0.67, 1. for primary $\gamma$, p, Ne, Fe respectively.}
\label{fig10}
\end{center}
\end{figure}
On the other hand, it is faster and more fault-tolerant and in 
that sense is expected to yield better results than other 
multivariate analysis techniques presently in use
for present exploratory exercise. We have used the jetnet 3.0 ANN
package developed by L\"onnblad et al. \cite{lblad}.
The transfer function used is of the sigmoid type $\sim$ ( 1 + exp
$(-2n))^{-1}$.  
Two hidden layers were used with backpropagation mode in 
this exercise alongwith optimized learning parameters.  
A sample of 12000 events was used during the training session 
of the network and the output values demanded are 0.0 for 
$\gamma$-rays, 0.33 for protons, 0.67 for Neon nuclei and 1.0 for 
Iron nuclei.
The test data set comprises
a total of 24000 events, belonging to $\gamma$-ray, proton, Neon and
Iron species, all 4 in an equal proportion.
First, the network was trained with only Hillas parameters, 
viz., L, W, A, and $\alpha$.
There is a clear distinction between $\gamma$ and hadron induced 
showers.
However, among hadrons, proton and the Neon distributions 
significantly overlap and the Iron component is 
relatively better separated as shown in
Fig.\ref{fig10}~. \\
In the second exercise, two fractal ($D_2$ and $D_6$) and two 
wavelet ($\beta_2$ and $\beta_6$) parameter values
have been used as inputs to the net and it is clear 
from Fig.\ref{fig10} that $\gamma$-rays and hadrons are 
again well separated, but
among hadrons, Neon and proton distributions 
are less overlapped while Iron is very well separated.
In the final ANN approach, Hillas parameters, fractal parameters
(two) and wavelet parameters (two) are used in tandem in the
input, again each with 12000 events of training and 24000
events of testing. 
The results are illustrated in Fig.\ref{fig10} 
where progenitor species, particularly $\gamma$-rays 
and Fe nuclei prove to be better separated from each other and also 
from proton and neon events compared with the procedures based 
on Hillas or fractal/wavelet parameters alone. 
Similar results are obtained by using 
different harder and softer cuts on the number of PE for the
training and generation samples. This largely precludes systematic
effects which could be due to the small size of the 
shower sample simulated by CORSIKA. 
With increasing number of parameters an increasing
number (squared) of simulated events for the training is required, 
thus for the combined training (Fig.\ref{fig10} upper panel)
the present number of simulated events seem to be small. 
An iterative way of using neural nets is therefore promising, i.e.
in a first step a separation of $\gamma$ to all hadrons
and, afterwards, a neural net analysis for the separation of
different charges of the hadronic cosmic rays. \\
Addressing finally the question what are the merits of the fractal
approach in analysing the images, as compared to the standard
procedures, one has to specify the detailed 
case and the aim of the analysis.
For example, if the analysis intends to prepare high purity samples
of $\gamma$-events at the expense of efficiency or a 
classification of all registered events in different primary 
mass groups with a reliable mapping of the mass composition, varying
strategies have to be employed and the optimum method has to be
dedicatedly explored.
 
\section{Discussion}
The possibility of using the TACTIC telescope to study the mass
composition of cosmic rays in UHE region has motivated us to look for non-traditional
approaches for parameterization of the Cerenkov image.
Proton is the main cosmic ray component which bombards
the upper layer of the atmosphere.  $\gamma$-ACE have comparatively 
lesser transverse development all through their passage in the longitudinal
direction  as compared with the corresponding situation in 
hadronic showers. The result is that Cerenkov light-producing 
charged components (e/$\mu$) are carried more closer to the 
ground level for hadronic showers and there
is intense pool of Cerenkov light close to the shower axis.
The electromagnetic processes in the hadronic showers are 
sustained by neutral-pion decay through 
$\pi^\circ \rightarrow \gamma + \gamma$. On the other hand, muons
resulting from decay of $\pi^\pm$ may have large transverse 
momenta imparted to them.  
These muons, on reaching the ground level,
produce local peaks in the overall Cerenkov lateral distribution 
profile. For ACE due to Neon and Iron nuclei, there is a large 
content of these muons and consequently Cerenkov image structure 
may be more diffuse than that in the case of protons.  
Hillas parameters are significantly sensitive to
distinguish $\gamma$-rays from the overall family of hadrons, 
but, as shown here, not good enough for efficient segregation 
of various Z groups in the cosmic-ray beam.
One reason for multifractal moments
to be effective as classifiers is because, as shown here, 
Cerenkov images are multidimensional patterns and therefore they can
be analyzed and parametrized by common pattern recognition methods
like fractals or wavelets. The resulting parameters are sensitive to
different physical features of the images, i.e. the shower 
development in the atmosphere.
On the other hand, the reason why wavelet moments
also seem to have an excellent potential as additional research tool
is that they search for differences in photon density 
gradients of the Cerenkov image.
The image profile gets more diffuse when we go from proton 
onwards to the iron component as the muon content also 
increases accordingly. 
The effect of the muon content on the Cerenkov image is to
diffuse the structure. It seems to be properly taken into account
by fractal and wavelet moments.
 
\section{Conclusions}
It has been shown here that Cerenkov images have a fractal
structure.  As a first follow-up of this important realization, it is
also indicated that multifractal dimensions and wavelet moments can
be used alongwith Hillas parameters to discriminate more efficiently
amongst gamma-rays, protons, Neon and Iron progenitor species
through the use of a properly-trained artificial neural net.
Multifractal and wavelet approach for analyzing Cerenkov images has
been discussed for the first time in gamma-ray astronomy
and this (preliminary) work suggests that the
outlook for using the resulting parameters as supplementary
classifiers for cosmic-ray mass-composition studies in the UHE
bracket is quite encouraging.

{\ack The work is embedded in the Indian-German bilateral agreement
of scientific-technical cooperation (WTZ INI-205).}

\end{document}